\documentclass[12pt, anon]{l4dc2024}

\usepackage{wrapfig}

\usepackage[subtle]{savetrees}

\usepackage[frozencache,cachedir=.]{minted}
\definecolor{codebg}{gray}{0.9}

\usepackage[capitalise]{cleveref}
\usepackage{lipsum}

\usepackage{xcolor}

\usepackage{xspace}
\newcommand{\lc}{\textit{L4CasADi}\xspace}

\usepackage{bm}
\usepackage{comment}
\usepackage{layouts}

\makeatletter
\newcommand*{\rom}[1]{\expandafter\@slowromancap\romannumeral #1@}
\makeatother

\title[L4CasADi]{Learning for CasADi \\ Data-driven Models in Numerical Optimization}

  \author{\Name{Tim Salzmann} \Email{Tim.Salzmann@tum.de}\\
   \Name{Jon Arrizabalaga} \Email{Jon.Arrizabalaga@tum.de}\\
   \addr Munich Institute of Robotics and Machine Intelligence, Technical University of Munich, Germany \\
   \Name{Joel Andersson} \Email{joel@jaeandersson.com}\\
   \addr Freelance software developer and consultant, Madison (WI), United States\\
   \Name{Marco Pavone} \Email{pavone@stanford.edu}\\
   \addr Stanford University, Stanford (CA), United States\\
   \and\Name{Markus Ryll} \Email{Markus.Ryll@tum.de}\\
   \addr Munich Institute of Robotics and Machine Intelligence, Technical University of Munich, Germany \\
   }

\begin{document}
\maketitle

\begin{abstract}
While real-world problems are often challenging to analyze analytically, deep learning excels in modeling complex processes from data.
Existing optimization frameworks like CasADi facilitate seamless usage of solvers but face challenges when integrating learned process models into numerical optimizations.
To address this gap, we present the \textit{Learning for CasADi} (L4CasADi) framework, enabling the seamless integration of PyTorch-learned models with CasADi for efficient and potentially hardware-accelerated numerical optimization.
The applicability of \lc is demonstrated with two tutorial examples: First, we optimize a fish's trajectory in a turbulent river for energy efficiency where the turbulent flow is represented by a PyTorch model.
Second, we demonstrate how an implicit Neural Radiance Field environment representation can be easily leveraged for optimal control with \lc.

\vspace{1em}

\noindent L4CasADi, along with examples and documentation, is available under MIT license at 
\vspace{0.5em}

\hspace{7em} \url{https://github.com/Tim-Salzmann/l4casadi}

\vspace{1em}
\end{abstract}

\section{Introduction}
\label{sec:intro}
Accurate mathematical problem formulation is at the core of every numerical optimization procedure. While many real-world problems are hard to formulate analytically, data-driven methods, especially deep learning methods, thrive in modeling complex processes from data. Efficiently integrating learned process models into numerical optimizations is challenging, as data-driven models and numerical optimization come with their respective tools and characteristics:
Deep learning models, commonly constructed in PyTorch~\cite{Paszke2019PyTorch:Library}, TensorFlow~\cite{Abadi2016TensorFlow:Systems}, or JAX~\cite{Bradbury2018JAX:Programs}, leverage first-order optimization methods (backpropagation) in time-expensive offline training on a large amount of data. Solutions to complex problems can be learned from data without defining any prior structure on the task. Once trained, however, their inference is fast and can be further accelerated on dedicated hardware (GPUs).
Numerical optimization techniques, in contrast, employ second-order methods, primarily interior point and sequential quadratic programming, to formulate optimization problems, potentially subject to constraints. Solving these problems necessitates formulating a specific problem structure and proficiently selecting a suitable solver algorithm (IPOPT, SNOPT, qpOASES, OSQP, ECOS, etc.\footnote{\cite{Wachter2006OnProgramming, Gill2005SNOPT:Optimization, Ferreau2014QpOASES:Programming, Stellato2020OSQP:Programs, Domahidi2013ECOS:Systems}}). Aiming to remove the overhead of adapting the syntax for each solver, multiple frameworks have appeared allowing seamless usage across them. Within convex problems, CVXPY~\cite{Diamond2016CVXPY:Optimization}, is a widely used Python wrapper. On the other hand, Pyomo~\cite{Bynum2021PyomooptimizationPython}, AMPL~\cite{Fourer1989AMPL:Language}, or CasADi~\cite{Andersson2019CasADi:Control} allow for a wider range -- including non-convex -- of optimization problems and solvers, including commercial ones as MOSEK~\cite{MOSEKApS2023TheToolbox.} or Gurobi~\cite{GurobiOptimizationLLC2023GurobiManual}.

Thus, both PyTorch and CasADi have established themselves as prominent tools within their respective research domains, owing to their comprehensive functionalities, user-friendliness, and adaptability. However, the increasing importance of data-driven approaches in optimization poses a challenge for CasADi as it lacks native support for learned functions.

We seek to close this gap by presenting the \textit{Learning for CasADi} (L4CasADi) framework, which enables the seamless integration of learned PyTorch models with the numerical optimization framework CasADi~\cite{Andersson2019CasADi:Control}. \lc enables hardware acceleration for learned components in a CasADi optimization formulation and can generate such problems as plain C/C++ code for efficient solution. By bringing this functionality to CasADi, tools building upon CasADi (such as acados~\cite{Verschueren2021Acados:Control}), as well as the large number of products already using CasADi, and the numerical optimization community at large can benefit from the same data-driven modeling capabilities.

\section{L4CasADi - Syntax and Usage}\label{sec:syntax}
\lc was designed with three key desiderata in mind: (\rom{1}) Simplicity for the user, (\rom{2}) generalizability across PyTorch model architectures, and (\rom{3}) efficiency in runtime. We will provide an insight into how these desiderata manifest within the user experience of \lc.
 
Similarly to PyTorch, \lc models are initially constructed in Python. Defining an \lc model in Python given a pre-defined PyTorch model is as easy as

\begin{minted}[fontsize=\small, bgcolor=codebg]{python}
import l4casadi as l4c
# Construct L4CasADi Model from PyTorch Model
l4casadi_model = l4c.L4CasADi(
    pyTorch_model,
    device='cpu',       # Device in ['cpu', 'gpu', 'mps']
    name='l4casadi_f'   # Unique name 
)
\end{minted}

\noindent where the architecture of the PyTorch model is unrestricted and large models can be accelerated with dedicated hardware.
Once an \lc model is defined it can be employed in a variety of settings, as will be outlined in the following.

\subsection{Python}\label{sec:python_usage}
Once defined, an \lc model can be seamlessly integrated with a CasADi symbolic graph within CasADi's Python interface. On the first call, \lc will automatically generate C++ code and compile the \lc model for runtime efficiency.

\begin{minted}[fontsize=\small, bgcolor=codebg]{python}
# Use L4CasADi Model in CasADi Symbolic Graph
y: casadi.MX = l4casadi_model(x: casadi.MX)
\end{minted}

The resulting symbolic output variable can be included in any further CasADi operations. This seamless integration empowers the utilization of CasADi's extensive toolkit to formulate and solve optimization problems involving \lc models.

\begin{minted}[fontsize=\small, bgcolor=codebg]{python}
# Minimize the L4CasADi Model using IPOPT
nlp = {'x': x, 'f': y}
solver = casadi.nlpsol("solver", "ipopt", nlp)
sol = solver()
\end{minted}

\subsection{Standalone Application}\label{sec:standalone}
To use an \lc model outside of the defining Python routine, the \lc model can also be explicitly built into an \textit{L4CasADi Shared Library Function}.

\begin{minted}[fontsize=\small, bgcolor=codebg]{python}
# Build the L4CasADi Model as L4CasADi Shared Library Function
l4casadi_model.build(x: casadi.MX)  #  Creates libl4casadi_f.so
\end{minted}
The shared library can easily be integrated into other programming environments, e.g. Matlab and C/C++.

\paragraph{Matlab}
A built \lc model can be loaded and used in Matlab using CasADi's Matlab interface.

\begin{minted}[fontsize=\small, bgcolor=codebg]{matlab}
% Load L4CasADi Model from dynamic library
l4casadi_model = casadi.external('l4casadi_f', 'libl4casadi_f.so');
% Use loaded L4CasADi Model in CasADi symbolic Graph
y: casadi.MX = l4casadi_model(x: casadi.MX)
\end{minted}

\paragraph{C/C++}
A built (\textit{L4CasADi Shared Library Function}) can be dynamically loaded from any C/C++ program. Additionally, \lc provides the \lc model as auto-generated C++ code which can be included in any C/C++ source project.

\begin{minted}[fontsize=\small, bgcolor=codebg]{python}
# Generate the L4CasADi Model as C++ Source Files.
l4casadi_model.generate(x: casadi.MX)  #  Generates l4casadi_f.cpp
\end{minted}

\subsection{Na\"ive L4CasADi}\label{sec:naive}
While \lc was designed with efficiency in mind by internally leveraging torch's C++ interface (see \cref{sec:arch}), this can still result in overhead, which can be disproportionate for small, simple models. Thus, \lc additionally provides a \textit{NaiveL4CasADiModule} which directly recreates the PyTorch computational graph using CasADi operations and copies the weights --- leading to a pure C computational graph without context switches to torch. However, this approach is limited to a small predefined subset of PyTorch operations --- only Multi-Layer-Perceptron style models and inference on CPU are supported.

\begin{minted}[fontsize=\small, bgcolor=codebg]{python}
naive_mlp: l4c.NaiveL4CasADiModule = l4c.naive.MultiLayerPerceptron(
    in_features,
    hidden_features,
    out_features,
    hidden_layers,
    activation='Tanh'
) 
l4c_model = l4c.L4CasADi(naive_mlp) 
y: casadi.MX = l4c_model(x: casadi.MX) 
\end{minted}

\section{Tutorial Examples}
To showcase the practical application of the proposed framework and how it opens up new research avenues at ease, we present two illustrative case studies. First, we formulate a trajectory generator that finds the minimum energy path for a fish swimming upstream in a turbulent river. The second case study demonstrates how \lc facilitates the incorporation of cutting-edge computer vision models into optimization problems. In this case, we optimize a collision-free trajectory through an implicit environment representation given as a Neural Radiance Field (NeRF). These two case studies are distinguished by their simplicity and ease of comprehension, and thereby, serve as excellent templates for users to adapt to more intricate scenarios. To this end, the code associated with these examples is available alongside \lc\footnote{\url{https://github.com/Tim-Salzmann/l4casadi/examples}}.

\subsection{Fish Navigation in Turbulent Flow\hspace{12.5em}\href{https://colab.research.google.com/github/Tim-Salzmann/l4casadi/blob/main/examples/fish_turbulent_flow/Fish_Turbulent_Flow.ipynb}{\includegraphics[scale=0.8]{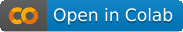}}}

The example shows how to design an optimization-based trajectory generator to navigate a turbulent fluid flow. The flow, and thus the fish's dynamics, are modeled by a Neural Network in PyTorch, while the optimal trajectory optimization problem is formulated in CasADi. By using \lc, both can be combined and optimized jointly. When doing so, we seek to find the minimum energy trajectory that allows a fish to navigate from the starting point to the goal. For this purpose, the fish needs to swim upstream a river where a circular stone causes the flow to be turbulent. 
\subsubsection{Problem Formulation and Implementation}
We model the fish as a planar point mass actuated by velocity commands under the influence of the river's velocity field:
\begin{equation}\label{eq:fish_ode}
    \dot{\bm{p}}(t) = \bm{v}(t)+ \bm{v}_{fl}(t,\bm{p}(t))\,,    
\end{equation}
where ${\bm{p}(t),\bm{v}(t)}\in R^2$ are the position and velocity of the fish at time $t$ and $\bm{v}_{fl}(t,\bm{p}(t))\in R^2$ is the velocity of the river at time $t$ and position $\bm{p}(t)$. In simpler terms, the fish's movement results from a combination of its own effort, and the influence of the turbulent flow. Eq.~\eqref{eq:fish_ode} can be written as a standard nonlinear dynamic system $f$, defined as
\begin{equation}\label{eq:nl_ode}
    \dot{\bm{x}}(t) = f(\bm{x}(t),\bm{u}(t),t) = \bm{u}(t)  + \bm{v}_{fl}(t, \bm{x}(t))\,,
\end{equation}
whose states and inputs are the fish's position $\bm{x}(t) = \bm{p}(t)$ and velocity command $\bm{u}(t)=\bm{v}(t)$. To add realism to the problem, we constrain the fish's actuation by imposing box constraints on the velocity commands $\bm{u}\in[\underline{\bm{u}}, \bar{\bm{u}}]$, limit the trajectory to lie within the rivers bounds $\left[\underline{\bm{p}},\bar{\bm{p}}\right]$ and constrain the trajectory to dodge the stone generating the turbulent flow.

To compute the trajectory that minimizes the fish's effort to reach the goal while swimming upstream, we formulate the following Nonlinear Program (NLP):
\begin{subequations}\label{ocp1}
	\begin{alignat}{3}
    \min_{\substack{\bm{x}_{\,0},\cdots,\,\bm{x}_{\,N},\\\bm{u}_{\,0},\cdots,\,\bm{u}_{\,N-1}}}&\sum_{k=0}^{N-2} \left|\left|\frac{\bm{u}_{k+1} - \bm{u}_k}{\Delta t}\right|\right|^2\label{ocp1:cost}\\
    \text{s.t.}\quad& \bm{x}_0=\bm{p}_{0}\,,\,\bm{x}_N=\bm{p}_{f}\,,\label{ocp1:start_end_states}\\
	&\bm{x}_{k+1} = \bm{x}_k + \Delta t \cdot f(\bm{x}_k,\bm{u}_k, t_k), &\quad&k = 0,\cdots,N-1\,, \label{ocp1:dynamics}\\
    &\underline{\bm{u}}\leq\bm{u}_k\leq\bar{\bm{u}}\,,    &\quad&k = 0,\cdots,N-1 \,,\label{ocp1:inputs}\\
    &\underline{\bm{p}}\leq\bm{x}_k\leq\bar{\bm{p}}\,,    &\quad&k = 0,\cdots,N \,,\label{ocp1:states1}\\
    &||\bm{x}_k||^2\geq r_{st}^2\,,    &\quad&k = 0,\cdots,N \,.\label{ocp1:states2}
\end{alignat}
\end{subequations}
where $\bm{p}_0$ and $\bm{p}_f$ are the initial and goal positions, $\Delta t$ is the time step and $r_{st}$ is the radius of the stone located at the origin. The cost function~\eqref{ocp1:cost} minimizes the differences in subsequent control (velocity) inputs which effectively minimizes the energy introduced into the system, constraint~\eqref{ocp1:start_end_states} sets the starting and ending positions, ~\eqref{ocp1:dynamics} enforces the system dynamics defined in~\eqref{eq:nl_ode}, ~\eqref{ocp1:inputs} bounds the input commands, ~\eqref{ocp1:states1} guarantees that the fish remains within the river's bounds and~\eqref{ocp1:states2} ensures that the fish does not collide with the stone. 
Notice that, due to the spatial bounds in~\eqref{ocp1:states2} and yet-to-be-defined turbulent flow model $\bm{v}_{fl}(\cdot)$ within the dynamic model $f(\bm{x},\bm{u})$ in~\eqref{ocp1:dynamics}, the NLP~\eqref{ocp1} is nonlinear and non-convex.

\lc allows for implementing the NLP in~\eqref{ocp1} in CasADi, modeling the turbulent flow $\bm{v}_{fl}(t,\bm{p}(t))$ by a Neural Network that has previously been trained in PyTorch.
Given the nonlinear and non-convex structure of~\eqref{ocp1}, the state-of-the-art interior-point solver IPOPT \cite{Wachter2006OnProgramming} is chosen as a solver.

\subsubsection{Outcome and Visualization}
\begin{figure}[h!]
    \centering
    \includegraphics[width=\linewidth]{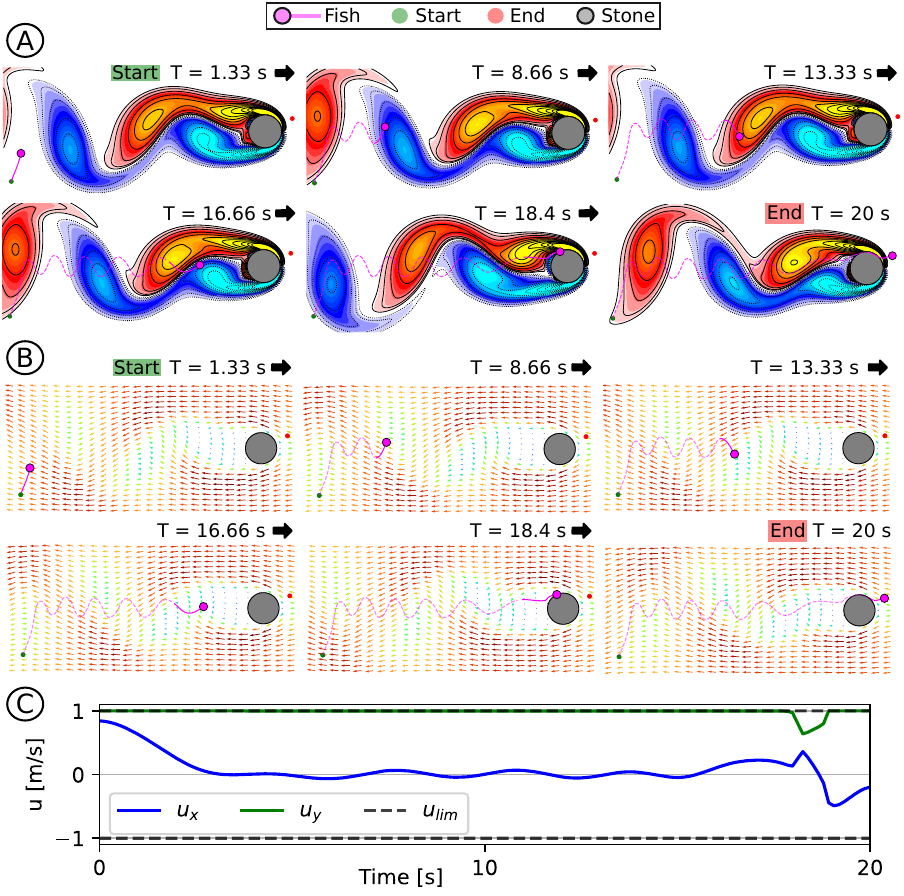}
    \vspace{-1em}
    \caption{Minimum energy trajectory (magenta) to navigate from the start (green) to the goal (red) in the presence of a turbulent flow represented by its vorticity (A) and velocity field (B). The trajectory generator is implemented in CasADi, while the flow is modeled by a Neural Network in PyTorch$^\text{\ref{note1}}$.}
    \label{fig:fish}
\end{figure}
The trajectory obtained from solving the NLP~\eqref{ocp1} is illustrated in Fig.~\ref{fig:fish} through a series of image sequences. The upper segment of figure (A) displays the magenta trajectory of the fish juxtaposed with the vorticity of the turbulent flow. Meanwhile, the middle segment (B) portrays the velocity field of the turbulent flow. In both instances, the stone is represented by a gray circle. The third segment (C) shows the velocity commands necessary to trace this trajectory. For a better insight, we encourage readers to view the accompanying animations for this example\footnote{\label{note1}Further visualizations and animations for the fish example can be found at \url{https://github.com/Tim-Salzmann/l4casadi/blob/main/examples/fish_turbulent_flow/README.md}.}.

Fig.~\ref{fig:fish} reveals that the computed trajectory strategically utilizes the velocity field of the turbulent flow to minimize the energy required for reaching the objective. To put it simply, the object behaves like it's riding/surfing the river's current. Examining Fig.~\ref{fig:fish} enables us to comprehend this phenomenon from three distinct perspectives: Firstly, in (A), it becomes evident that the fish exploits the positive vortices (blue) to ascend and the negative ones (red) to descend. This observation is similarly evident in (B), where the fish adeptly positions itself per the velocity field, enabling it to exert minimal effort in the y-axis direction throughout the majority of the navigation, as depicted by the green line in (C).


\clearpage
\subsection{Collision-free Trajectory Optimization in NeRFs
\hspace{6.4em}\href{https://colab.research.google.com/github/Tim-Salzmann/l4casadi/blob/main/examples/nerf_trajectory_optimization/NeRF_Trajectory_Optimization.ipynb}{\includegraphics[scale=0.8]{figures/open_in_colab.pdf}}}

Neural Radiance Fields (NeRFs) are a powerful 3D representation technique that leverages deep learning to reconstruct 3D scenes from a collection of 2D images. NeRFs encode the scene's geometry and appearance into a continuous function, allowing for the rendering of photorealistic images from arbitrary viewpoints. As a byproduct, the densities (translucencies) of the scene's objects at any point in the 3D environment are implicitly captured in the network. 
In this example, we demonstrate that such state-of-the-art learned models from computer vision research can be easily incorporated into optimization procedures using \lc. To showcase this, we present the problem of finding a collision-free trajectory through the densities represented by a learned NeRF, where densities below a predefined threshold are deemed as unobstructed regions within the environment.

\subsubsection{Problem Formulation}
The trajectory to be planned is assumed to be given by a time-parametric polynomial of degree $9$:
\begin{equation}\label{eq:nerf_traj}
    \bm{r}(\bm{c},t) = \sum^9_{i=0} \bm{c}_i t^i\,
\end{equation}
where the parametric variable $t$ is time and $\bm{c}\in\mathbb{R}^{9\times3}$ are the polynomial's coefficients.

\lc enables the integration of a NeRF as an implicit environment representation into the optimization problem. For this purpose, we define the following function:
\begin{equation}\label{eq:nerf_occ}
\rho=f_{\text{NeRF}}(\bm{p})\,
\end{equation}
where $\rho\in\mathbb{R}^+$ is the density of a location whose Euclidean coordinates are given by $\bm{p}\in\mathbb{R}^3$, i.e., it returns $0$ for obstacle-free space while its value increases as the location becomes occupied.

Having defined the analytical expression of the trajectory in~\eqref{eq:nerf_traj} and the NeRF-based environment representation in~\eqref{eq:nerf_occ}, we formulate a NLP that minimizes the curve's snap, while ensuring that it remains collision-free:
\begin{subequations}\label{ocp2}
	\begin{alignat}{3}
    \min_{\bm{c}}\sum_{k=0}^{N}& \left|\left|\bm{r}^{(4)}(\bm{c},t_k)\right|\right|^2\label{ocp2:cost}\\
    \text{s.t.}\quad& \bm{r}(\bm{c},0)=\bm{p}_{0}\,,\,\bm{r}^{(1)}(\bm{c},0)=\bm{0},\,\bm{r}^{(2)}(\bm{c},0)=\bm{0}\,,\label{ocp2:start}\\
    & \bm{r}(\bm{c},T)=\bm{p}_f\,,\,\bm{r}^{(1)}(\bm{c},T)=\bm{0},\,\bm{r}^{(2)}(\bm{c},T)=\bm{0}\,,\label{ocp2:end}\\
    &\bar{\rho}>f_{\text{NeRF}}\left(\bm{r}(\bm{c},t_k)\right)\,,    &\quad&k = 0,\cdots,N \,,\label{ocp2:nerf}
\end{alignat}
where $r^{(n)}(\cdot)$ is the $n$-th time derivative of eq.~\eqref{eq:nerf_traj}, $N$ is the number of evaluation points, $T$ is the total time assigned to the trajectory. Constraints~\eqref{ocp2:start} and~\eqref{ocp2:end} define the starting and ending conditions
and~\eqref{ocp2:nerf} ensures that the NeRF density at all points along the trajectory is below a threshold $\bar{\rho}$ and thus collision-free.
\end{subequations}

This problem, in its original form without the NeRF, is well-studied in the planning and trajectory optimization community. Its mathematical structure allows it to be formulated as a quadratic programming (QP) problem, which can be solved efficiently.
Being able to incorporate computer vision models in the problem formulation expands the scope of research possibilities but can simultaneously introduce greater complexity to the optimization task: The NeRF in~\eqref{eq:nerf_occ} makes the resulting NLP highly non-convex due to the non-smooth density landscape as depicted in \cref{fig:nerf2d}.

\begin{figure}
\centering
\subfigure[]{\includegraphics[width = 0.3\textwidth]{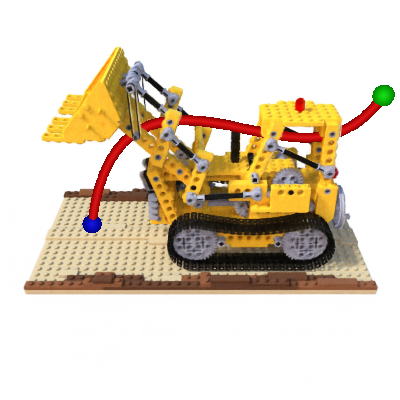}}
\subfigure[]{\includegraphics[width = 0.3\textwidth]{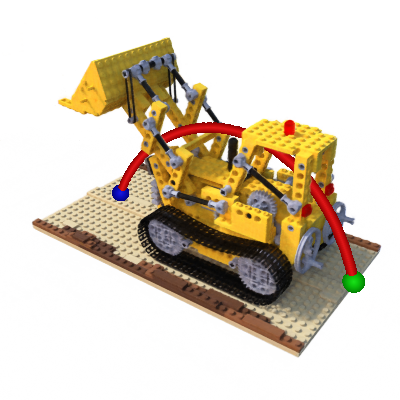}}
\subfigure[]{\includegraphics[width = 0.3\textwidth]{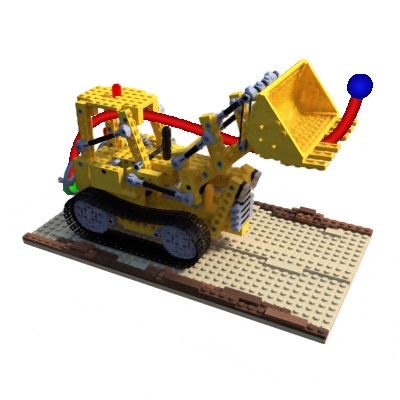}}
\caption{NeRF renders of three examples with a collision-free minimum snap optimized trajectory through the NeRF object. In each example, we vary the start point (blue) and goal point (green).}
\label{fig:nerf}
\vspace{-2em}
\end{figure}

\subsubsection{Implementation}
We utilize a pre-trained NeRF of a yellow Lego bulldozer, a commonly employed example in NeRF research. IPOPT~\cite{Wachter2006OnProgramming} is again used as a solver. However, due to the complexity of the problem, we apply a two-stage approach. The first phase optimizes the trajectory to closely follow a sparse set of predetermined collision-free points, excluding the NeRF constraint in~\cref{ocp2:nerf}. This results in a sub-optimal yet feasible trajectory. In the second phase, we initialize the solver with the first-phase solution and optimize for the full NLP in~\cref{ocp2} including constraining all points to have a lower NeRF density than $\bar{\rho} = 1$. To ensure IPOPT utilizes the warmstarted trajectory in phase two, we initialize IPOPT with a small barrier parameter $\mu = 1e^{-4}$ (See Eq. (3a) in~\cite{Wachter2006OnProgramming}). This compels IPOPT to remain within the region of feasible solutions from the beginning.

\subsubsection{Outcome and Visualization}
\begin{wrapfigure}{r}{0.4\textwidth}
 \vspace{-4.5em} 
  \begin{center}
    \includegraphics[trim={0 0 1.0cm 0}, clip, width=0.38\textwidth]{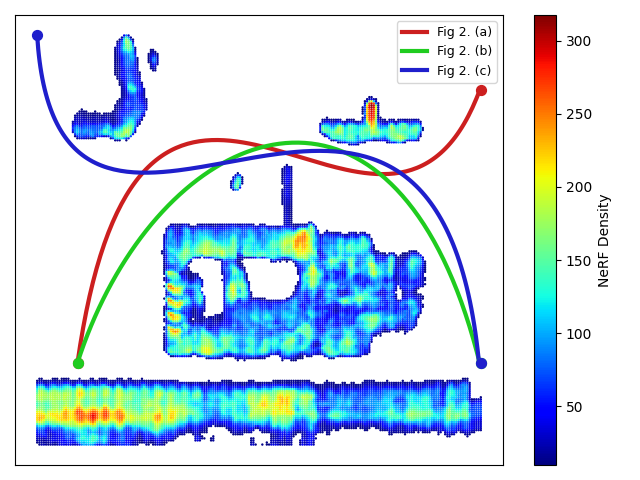}
  \end{center}
  \vspace{-2em}
  \caption{2D slice of \cref{fig:nerf}.}
  \label{fig:nerf2d}
\end{wrapfigure}
The optimized trajectory for three different configurations with varying start- and goal-points are shown in~\cref{fig:nerf}. For a better understanding of the problem space and the solution trajectories, we also visualize a 2D slice through the environment with the three optimal trajectories in~\cref{fig:nerf2d}. The optimal minimal snap trajectories are clearly collision-free with the NeRF's density representation. We encourage the reader to view the accompanying animations for this example\footnote{Further visualizations and animations for the NeRF example can be found at \url{https://github.com/Tim-Salzmann/l4casadi/blob/main/examples/nerf_trajectory_optimization/README.md}.}.

\section{Framework Architecture}\label{sec:arch}

\begin{figure}
    \centering
    \includegraphics[width=\linewidth]{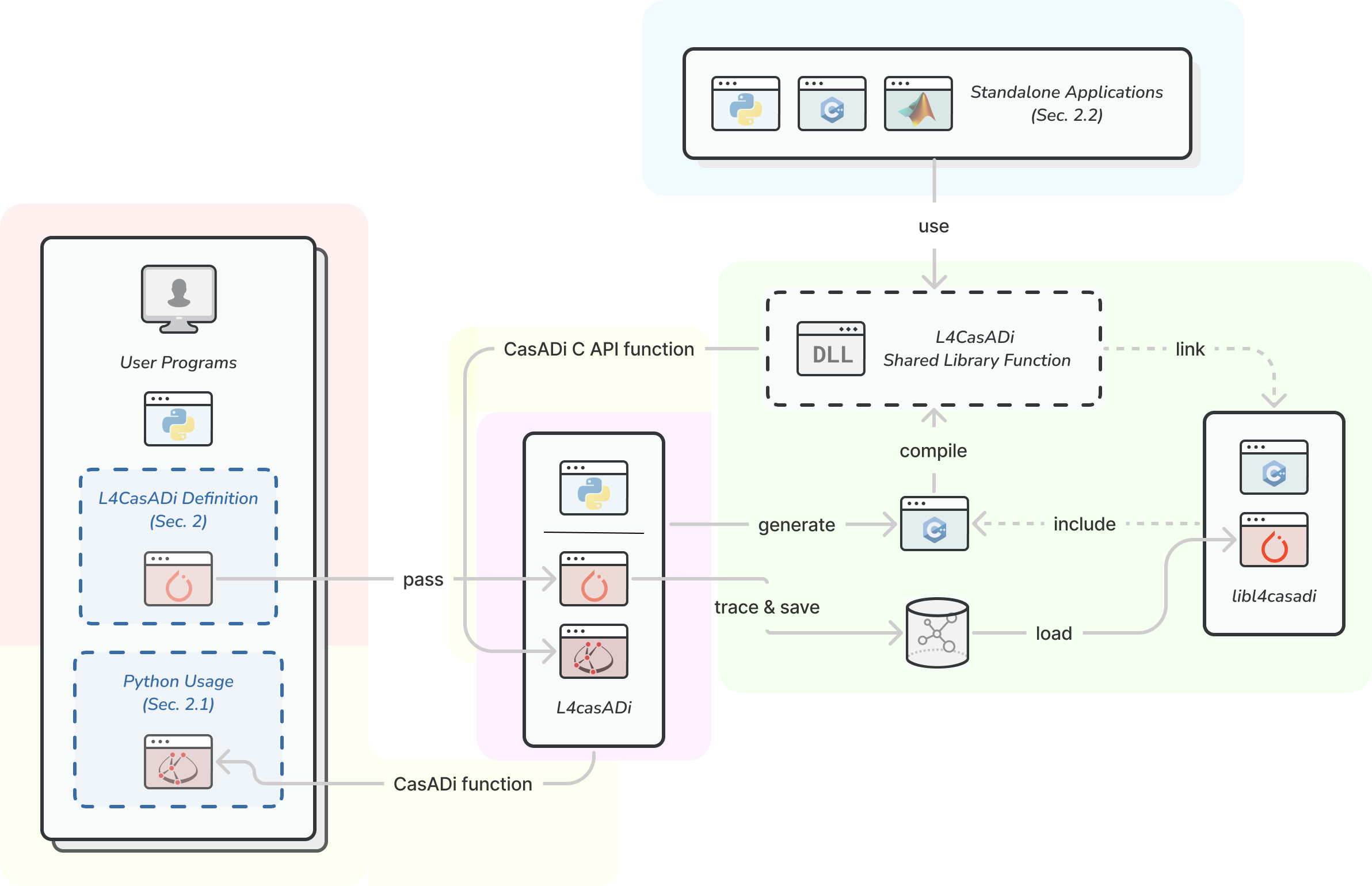}
    \caption{Software Architecture of \lc. The block details are presented in Section \ref{sec:arch}.}
    \label{fig:arch}
\end{figure}

To facilitate a comprehensive grasp of the \lc framework's internal operations, the software architecture of the \lc framework is presented in~\cref{fig:arch}. We will guide the reader through the architecture step-by-step, using the figure's background colors for orientation.

\paragraph{Red}
In accordance with~\cref{sec:syntax}, \lc models are invariably constructed in a Python user program. This entails constructing the PyTorch model and subsequently passing it to the L4CasADi Python class instantiation.

\paragraph{Purple}
\lc symbolically traces the forward pass, the Jacobian, and Hessian of the passed PyTorch model and stores these symbolic representations. Tracing is supported for virtually every PyTorch model architecture. Further, \lc auto-generates the function as C++ code.

\paragraph{Green}
The generated code includes the functionality of the \textit{libl4casadi} library, which facilitates the evaluation of saved PyTorch Traces efficiently within C/C++. These functionalities are exported adhering to the CasADi C interface. The generated code is automatically compiled to a shared library (\lc Shared Library Function).

\paragraph{Yellow}
For direct use of the \lc model within Python (as described in \cref{sec:python_usage}) the shared library, exporting function symbols adhering to CasADi's C interface, is loaded by the Python interface of CasADi within the \lc Python class (purple). L4CasADi returns this external function to the Python user program as a CasADi function, which at that point can be seamlessly included in any CasADi computational graph and optimization. By providing derivative information complying with CasADi's C interface naming convention, first and second-order derivatives of the L4CasADi model are available in the CasADi graph.

\paragraph{Blue}
In addition to directly using the L4CasADi model in the instantiating Python program, the compiled \lc Shared Library Function can be loaded and used in C/C++, Matlab, or Python standalone applications (See \cref{sec:standalone}).

\section{Related Work}
The increasing prevalence of data-driven models in optimization has spurred a surge of scholarly contributions, making it infeasible to comprehensively review the entire landscape of this domain within the scope of this discussion. Instead, we will concentrate on specific frameworks that integrate learned models with numerical optimization techniques, specifically PyTorch and CasADi.

\paragraph{Learning and Numerical Optimization}
Multiple approaches have been proposed to bring advanced numerical optimization algorithms within learning frameworks like PyTorch and TensorFlow \cite{Amos2018DifferentiableControl, Wang2023GOPS:Applications}. However, these integrations often result in the re-implementation of individual solver algorithms within the learning framework, which may not exhibit the same robustness and maturity as the well-established solvers within CasADi.
Recently, NeuroMANCER~\cite{Drgona2023NeuroMANCER:Regularizations} enables the formulation of optimization problems entirely in PyTorch. Naturally, learned components in PyTorch can be included, even optimized within the problem solution. However, because NeuroMANCER is restricted to the PyTorch environment they are mostly limited to the first-order solvers within --- not using second-order approaches such as IPOPT or SQP which have been proven to be efficient and robust. 

\paragraph{PyTorch and CasADi}
The demand for such a framework within the community is evident from prior endeavors aimed at uniting the two concepts of learning in PyTorch and optimization in CasADi. Multiple packages, \textit{do-mpc}~\cite{Fiedler2023Do-mpc:Control}, \textit{HILO-MPC}~\cite{Pohlodek2022FlexibleHILO-MPC}, and \textit{ML-CasADi}~\cite{Salzmann2023Real-TimePlatforms}, provide the capabilities to rebuild simple architectures of PyTorch models directly in CasADi by copying the learned weight tensors and formulate matrix multiplications and activation functions in CasADi. This approach, similar to Na\"ive \lc (\cref{sec:naive}), is restricted to models comprised of the limited CasADi function set. Further, CasADi's function set is not optimized for large matrix multiplications and can not use hardware acceleration.
Additionally, \cite{Salzmann2023Real-TimePlatforms} calculates function evaluations and sensitivities for the learned model separated from CasADi in the PyTorch framework and subsequently injects these results into the CasADi graph. 
The applicability of this approach, however, is limited to specific optimization algorithms and introduces inefficiencies in the form of excessive context switches and memory transfers between CasADi and PyTorch.

\newpage
\bibliography{references}

\end{document}